\documentclass[aps,pre,reprint,groupedaddress]{revtex4-1}
\usepackage[dvipdfmx]{graphicx}
\usepackage{subfigure}
\usepackage[dvipdfmx]{color}
\usepackage{amssymb}

\begin{document}

\title{Temporal analysis of acoustic emission from a plunged granular bed}
\author{Daisuke Tsuji and Hiroaki Katsuragi}
\affiliation{Department of Earth and Environmental Sciences, Nagoya University, Furocho, Chikusa, Nagoya 464-8601, Japan}
\date{\today}

\begin{abstract}
The statistical property of acoustic emission (AE) events from a plunged granular bed is analyzed by means of actual time and natural time analyses. These temporal analysis methods allow us to investigate the details of AE events that follow a power-law distribution. In the actual time analysis, the calm time distribution and the decay of the event-occurrence density after the largest event (i.e., Omori-Utsu law) are measured. Although the former always shows a power-law form, the latter does not always obey a power law. Markovianity of the event-occurrence process is also verified using a scaling law by assuming that both of them exhibit power laws. We find that the effective shear strain rate is a key parameter to classify the emergence rate of power-law nature and Markovianity in the granular AE events. For the natural time analysis, the existence of self organized critical (SOC) states is revealed by calculating the variance of natural time $\chi_k$, where $k$th natural time of N events is defined as $\chi_k=k/N$. In addition, the energy difference distribution can be fitted by a $q$-Gaussian form, which is also consistent with the criticality of the system. 
\end{abstract}


\maketitle

\section{Introduction}
\label{sec:sec1}
Power-law distributions can be observed in various fields of natural and artificial phenomena~\cite{Clauset2009}. Examples include the distribution of incomes (Pareto's law), the appearance frequency of English words (Zipf's law), the number of meteorites impacting on a planet, and the number of species per genus in flowering plants~\cite{Reed2001}. For seismic activity, one of the best known power-law relations is Gutenberg-Richter (GR) law~\cite{GR-law}:
\begin{equation}
G(Q)\sim Q^{-\gamma},
\label{eq:size_power_law}
\end{equation}
where $Q$, $G(Q)$, and $\gamma$ are the emitted energy per event, its occurrence frequency, and a characteristic exponent (positive constant), respectively. Recently, experiments and simulations of soft materials which are related to seismic activity have been performed. These studies have reported various power-law event-size distributions (e.g., sliding friction of gels~\cite{Yamaguchi2011} and granular avalanches in simulations~\cite{Hatano2015}). Although this kind of power-law event-size distributions has also been discovered in many natural phenomena (e.g., forest fire areas~\cite{Malamud1998}, floods~\cite{Turcotte1993}, fragments~\cite{Astrom2006}, and Tsunami runup heights~\cite{Burroughs2005}), they are empirical laws and not fully understood in terms of their physical origin. 

The physical mechanisms determining the power-law exponent have long been studied. For instance, the exponent of power-law size distribution in brittle fragmentation shows particular relations to the higher order moment and system size~\cite{Katsuragi2003,Katsuragi2004,Katsuragi2005}. Besides, the exponent value depends on the dimensionality and the way of crack propagation~\cite{Astrom2006}. Also, stress relaxation mechanisms in plastic deformation can be classified by the power-law exponent of stress drop distributions~\cite{Niiyama2015}. Although some fundamental aspects concerning the power-law size distributions have been revealed as aforementioned, much deeper investigation associated with the appropriate classification of the system is necessary for truly understanding the universality of the whole power-law distributions.

In this study, acoustic emission (AE) burst events emitted from a plunged granular bed are particularly examined. In Ref.~\cite{Matsuyama2014}, the behavior of granular matter has been studied using AE technique, in which the size distribution of AE burst events obeys power law. In the experiment, the power-law exponent ($\gamma$ in Eq.~(\ref{eq:size_power_law})) varies depending on experimental conditions. Although the exponent value was  related to the mode of deformation (brittle-like or plastic-like), the analysis has still been very qualitative in Ref.~\cite{Matsuyama2014}. Thus, further detailed analyses are necessary to identify the underlying physical mechanisms governing the power-law nature of granular AE events. Such a deeper understanding of granular AE events might also provide a universal framework for various power-law distributions. Moreover, because the statistical behavior of dry granular matter is somewhat similar to that of seismicity~\cite{Duran2000}, the detailed study of granular behavior could be helpful to understand geophysical phenomena as well. 

The most serious deficiency in the previous analyses of AE event-size distributions is a lack of temporal information. The power-law event-size distribution such as GR law usually neglects the time series of event occurrence. It only deals with the size of events whilst the events indeed occur in the time series. In order to consider the temporal information, here we employ two analysis methods: actual time and natural time analyses. 

In the actual time analysis, the amplitude of events is omitted in contrast to the analysis of event-size distributions such as GR law. Then, the time interval between successive AE events called {\it calm time} (a.k.a.~interoccurrence time or waiting time) and the event-occurrence density are measured. The distribution of calm time has been found to be power law in many AE measurements (e.g., tensile failure experiment of paper sheets~\cite{Salminen2002} and volcanic rocks at Stromboli~\cite{Diodati1991}). For the event-occurrence density, the power-law decay of aftershock activity is known as Omori-Utsu (OU) law in seismology~\cite{Omori1894,Utsu1961}. OU-like behavior is also observed in various AE measurements (e.g.,~microfracturing in a compressed rock~\cite{Hirata1987,Ojala2004}). 

If both these quantities (calm time and event-occurrence density) exhibit power-law distributions and each power-law exponent can be determined, we can evaluate Markovianity of the event time series using these exponents. This powerful method to verify Markovianity was first developed for analyzing real seismic data, and non-Markov nature of the real seismic activity was revealed by Abe and Suzuki~\cite{Abe2009,Abe2012}. In the current study, we are going to discuss the statistical property of granular AE events by applying this method to the granular AE event data. 

The natural time analysis, on the other hand, discards the calm time information. It only uses the order of events and corresponding amplitudes. The idea of natural time has also been proposed for the analysis of seismicity~\cite{Varotsos2001, Varotsos2003, Varotsos2003(1), Varotsos2004, Varotsos2005, Varotsos2005(1), Varotsos2006, Varotsos2010, Varotsos2011, Sarlis2006, Sarlis2011}. For an event time series comprising $N$ events, natural time $\chi_k$ serves as an index for the occurrence of the $k$th event and is defined as $\chi_k=k/N$. Then the variance $\kappa_1$ is computed as
\begin{equation}
\kappa_1=\sum_{k=1}^{N}p_k\chi_k^2-\left(\sum_{k=1}^{N}p_k\chi_k\right)^2=\langle\chi^2\rangle-\langle\chi\rangle^2,
\label{eq:kappa_def}
\end{equation}
where $p_k=Q_k/\sum_{i=1}^{N}Q_i$ is the normalized released energy $Q_k$ in the $k$th event. The seismic electrical signals (SES) right before earthquakes tend to show a critical value $\kappa_1=0.07$~\cite{Varotsos2001}. Similar $\kappa_1$ behaviors can be confirmed in some self organized critical (SOC) systems~\cite{Varotsos2010, Varotsos2011, Sarlis2006, Sarlis2011}, where the original concept of SOC was introduced by Bak et al.~\cite{Bak1987}. Furthermore, the AE signals from the deformed rock have exhibited the behavior similar to seismicity in terms of the natural time analysis~\cite{Vallianatos2013}. 

In this study, the temporal properties of AE events emitted from a plunged granular bed are throughly investigated through these analyses. The relation between the measured results and the previously obtained event-size distributions (GR law)~\cite{Matsuyama2014} is also discussed on the basis of the experimental data.

\section{Experiment}
\label{sec:sec2}
The experimental methodology and data used in this study are the same as those in Ref.~\cite{Matsuyama2014}. Glass beads (grain diameter $d=0.4,~0.8,$ or $2.0$~mm) are poured into a cylindrical plexiglass container. A steel sphere (radius $r=5,~10,$ or $20$~mm) is then penetrated into the granular bed. The penetration speed is fixed as $v=0.5,~1.0,$ or $5.0$~mm~s$^{-1}$. The top surface of the granular bed is open to the atmosphere and any confining pressure is not applied to the bed. An AE sensor (NF AE-900s-WB) is buried and fixed in the granular bed to capture AE events created by the penetration. The AE sensor is a piezoelectric transducer which converts dynamic motions (e.g., ultrasonic elastic wave) into electric signals~\cite{Grosse2008}. Because the AE signals are very weak, they are amplified by an amplifier (NF AE-9913) and a discriminator (NF AE-9922). The sampling rate of the AE data is $1$ MHz. Three experimental realizations for each setup of experimental conditions are performed to check the reproducibility. In general, granular behaviors have strong memory effect and history dependence~\cite{Katsuragi2015}. The sphere penetration must remain its memory in the granular bed such as the force chain structure in the tapped granular bed~\cite{Iikawa2015}. Thus, the fresh granular bed is deposited before every experimental run to erase the memory of penetration.

\begin{figure}[!b]
\begin{center}
\includegraphics[width=86mm]{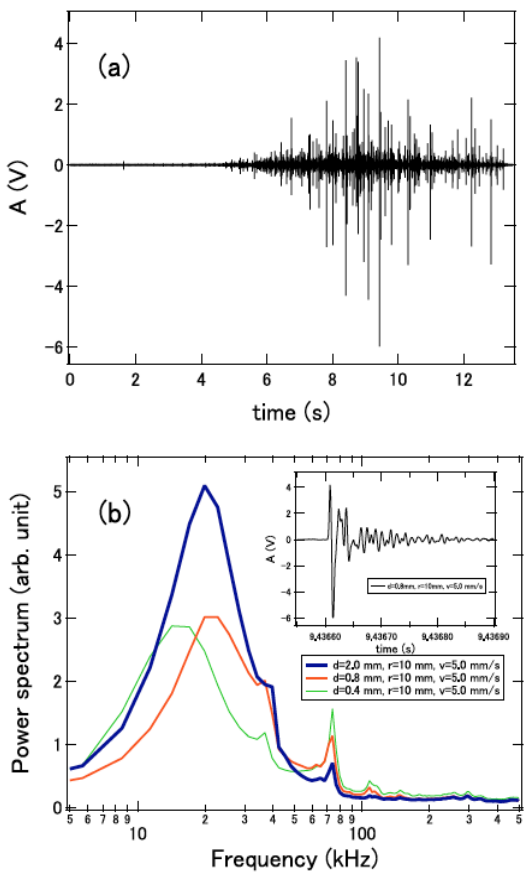}
\caption{(Color online) (a) Example of the AE signals during the penetration of a sphere into a glass bead bed. (b) Average power spectra of AE events produced per penetration for some experimental conditions. The inset shows an example of typical attenuating oscillation with short decay time.}
\label{fig:raw}
\end{center}
\end{figure}

Let us summarize the experimental results briefly. A raw data example of AE signals $A$ in volt as a function of time is shown in Fig.~\ref{fig:raw}(a). While the origin of time in Fig.~\ref{fig:raw}(a) is defined by $z=0$ ($z$ is the penetration depth of the intruder) as similar to Ref.~\cite{Matsuyama2014}, it is arbitrary in the current study. Actually, the origin of time $t=0$ will be defined later by the main event. The corresponding experimental conditions are $d=0.8$~mm, $r=10$~mm, and $v=5.0$~mm~s$^{-1}$. From now on, we use these experimental conditions for subsequent plots shown in this paper unless otherwise noted. Because all AE events show typical attenuating oscillation with short decay time as depicted in the inset of Fig.~\ref{fig:raw}(b), each AE event can be picked up from the measured AE signals using a threshold value $A_{th}=0.06$ V and deadtime $t_{dead}=300$ ${\mu}$s~\cite{Matsuyama2014}. The total number of AE events identified by this method ranges from $10^2$ to $10^4$. Figure~\ref{fig:raw}(b) shows average power spectra of AE events for several experimental conditions. The dominant frequencies, $20$ kHz and $75$ kHz, seem to be independent of the experimental conditions and might result from complex coupling of both the AE device and the granular media. For more details of the experimental setting, see Ref.~\cite{Matsuyama2014}.

\section{Actual time analysis}
\label{sec:sec3}
\subsection{Calm time distribution}
\begin{figure}[thbp]
\begin{center}
\includegraphics[width=86mm]{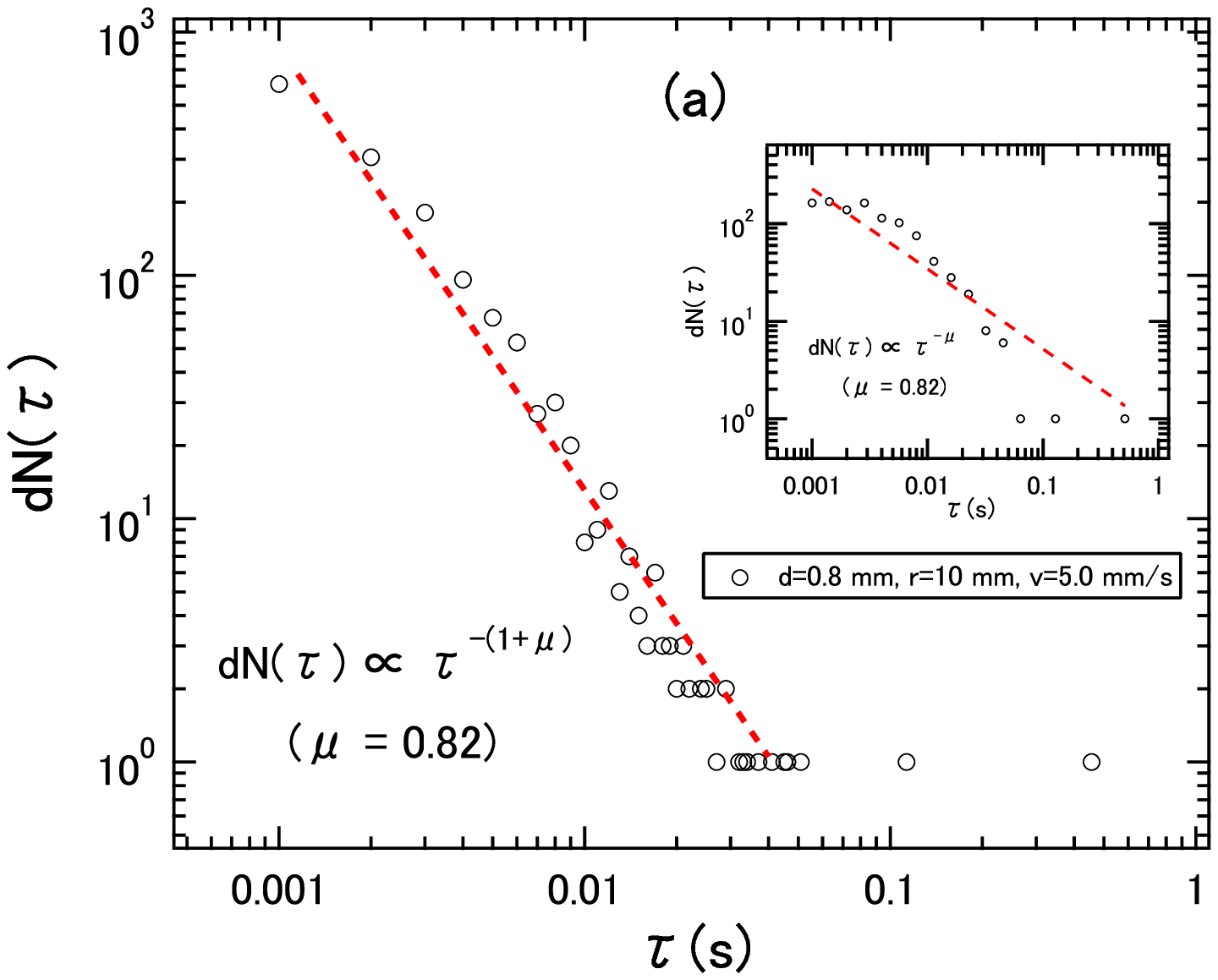}
\includegraphics[width=86mm]{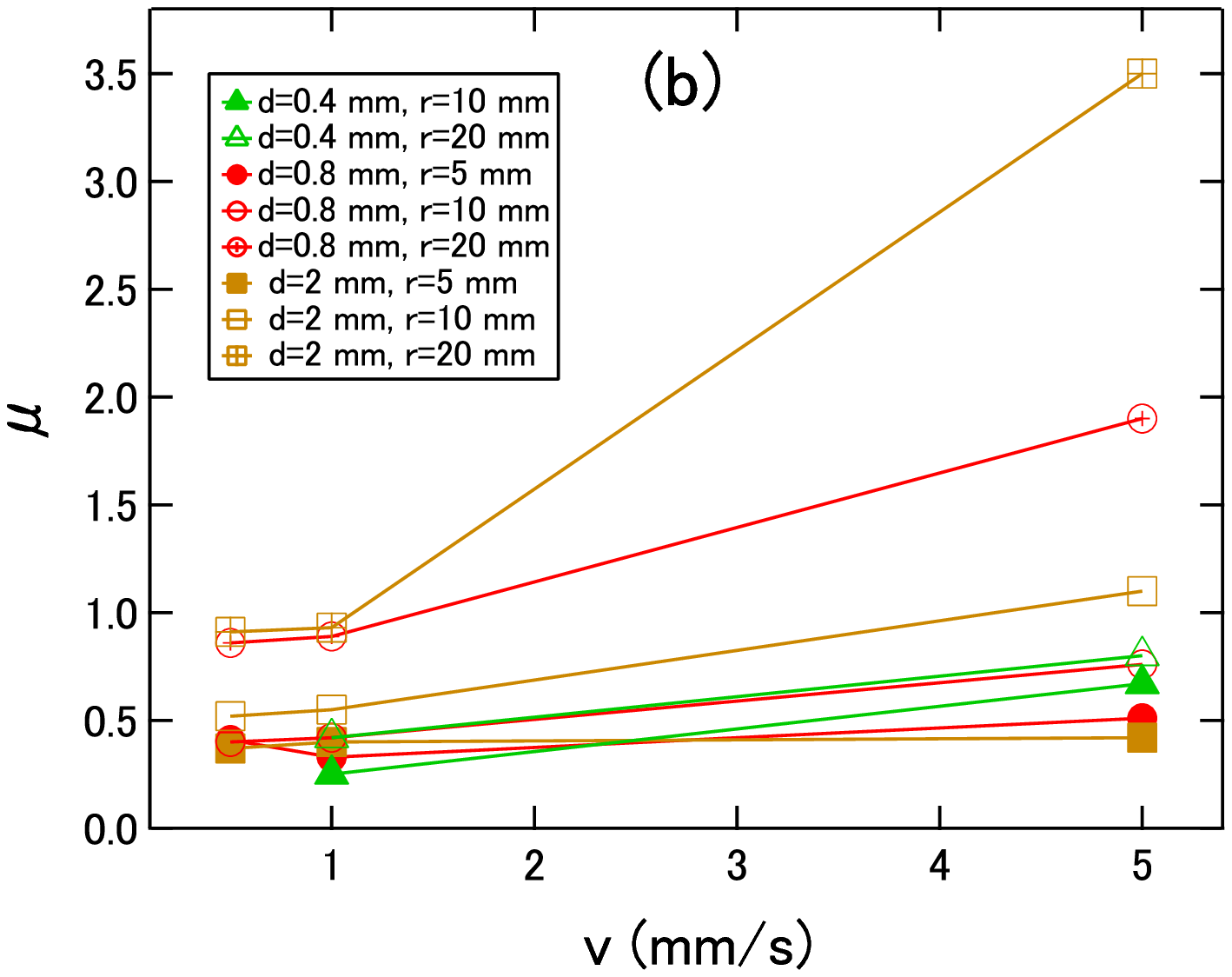}
\end{center}
\caption{(Color online) (a) Calm time distribution $dN(\tau)$ obeying a power law (Eq.~(\ref{eq:P_tau})). The inset shows $dN(\tau)$ with logarithmically increasing bins, where the slope corresponds to $-\mu$, not $-(1+\mu)$. The bin widths are given by $0.001(\sqrt{2})^n$~s (e.g., $0.001, 0.001\sqrt{2}, 0.002$ ...~s). Red broken lines in both plots show the power law with $\mu=0.82$ which is obtained by MLE method. (b) The exponent of the calm time distribution $\mu$ as a function of the penetration speed $v$. The $\mu$ value depends on various experimental conditions in contrast to $\gamma$ in Eq.~(\ref{eq:size_power_law}). The data of calm time used in this figure are obtained after the main event defined in Sec.~\ref{sec:OU}.}

\label{fig:calm_time_distribution}
\end{figure}

First, we focus on the analysis of calm time $\tau$ which corresponds to the time interval between two successive events. Specifically, the frequency distribution of calm time $P(\tau)$ is measured. The occurrence time of each event is determined by the moment at which the signal amplitude exceeds the threshold value $A_{th}$. In Fig.~\ref{fig:calm_time_distribution}(a), an example of $P(\tau)$ distribution obeying the power-law form is presented. All other $P(\tau)$ distributions also exhibit power-law forms. Namely, $P(\tau)$ always obeys 
\begin{equation}
P(\tau) = \frac{dN(\tau)}{N_{tot}}\sim\tau^{-(\mu+1)},
\label{eq:P_tau}
\end{equation}
where $dN(\tau)$ and $N_{tot}$ are the number of events with calm time $\tau$ intervals and the total number of events in the time series, respectively. The minimum (binning) timescale used in $dN(\tau)$ measurement is $1$ ms. Since the number of events is limited and the finite size effect must be considered carefully, here we employ the method of maximum likelihood estimation (MLE)~\cite{Goldstein2004,Newman2005,Clauset2009} to determine the exponent value. MLE determines the exponent to maximize the likelihood function $L$ defined as
\begin{equation}
L(1+\mu|\tau)=\prod_{k=1}^{N} \frac{\tau_k^{-(1+\mu)}}{\zeta(1+\mu)},
\label{eq:MLE}
\end{equation}
where $\tau_k$ is the calm time of the $k$th event normalized to the minimum value and $\zeta(1+\mu)$ is the Riemann zeta function expressed as $\sum_{k=1}^{N}k^{-(1+\mu)}$. To apply MLE, the exponent $\mu$ must be greater than zero. Otherwise, the Riemann zeta function diverges. Because MLE enables us to avoid large bias error in many frequency distributions (e.g., \cite{Goldstein2004, Bauke2007, White2008}), MLE is a preferred method for accurately estimating the power-law exponent. The determined $\mu$ value in the data of Fig.~\ref{fig:calm_time_distribution}(a) is $0.82$. To check the validity of the determined exponent from another aspect, the frequency distribution of $\tau$ with the logarithmic bins using a constant rate $\sqrt{2}$ is shown in the inset of Fig.~\ref{fig:calm_time_distribution}(a). Although the data in the inset are scattered, the global trend can be reproduced by the power-law fitting determined by MLE ($\mu=0.82$). 

For the obtained power-law exponent, $\mu$ depends on the experimental conditions. According to the previous study~\cite{Matsuyama2014}, the exponent of event-size distributions (GR law) $\gamma$ mainly depends on the grain size $d$. In contrast, the exponent of the calm time distribution $P(\tau)$ is sensitive to various parameters. Figure~\ref{fig:calm_time_distribution}(b) shows the $v$ dependence of the exponent $\mu$, where $v$ and $\mu$ are positively related. This means that the faster the penetration speed is, the more the relative frequency of shorter calm time increases. 

\subsection{Event-occurrence density distribution}
\label{sec:OU}
Second, the number of events occurring per unit time after the main event $S(t)$ is considered. $S(t)$ is defined by
\begin{equation}
S(t) = \frac{dN'(t)}{dt} \sim t^{-p},
\label{eq:S_t}
\end{equation}
where $p$ is a characteristic exponent, $t$ stands for the time elapsed after the main event, and $dN'(t)$ is the number of events occurring in the short time interval between $t$ and $t+dt$. We call $P(t)$ the {\it event-occurrence density}. Because the time $t=0$ should be defined by the moment of the main-event occurrence, it is necessary to locate the main event. In usual seismic activity, the main event (mainshock) can be located by subsequent smaller aftershocks~\cite{Lay1995}. Thus, in this work, we employ the largest AE event in the time series as a main event. Then, the event-occurrence density after the main event is measured from the experimental data. For real seismic activity, the universal power law is well known as OU law, which states that $S(t)$ decays with time following a power-law relationship expressed on the right-hand side in Eq.~(\ref{eq:S_t}). The exponent $p$ usually ranges in $0.8-1.5$ for real seismic activity~\cite{Varotsos2006}. 

\begin{figure}[thbp]
\begin{center}
\includegraphics[width=86mm]{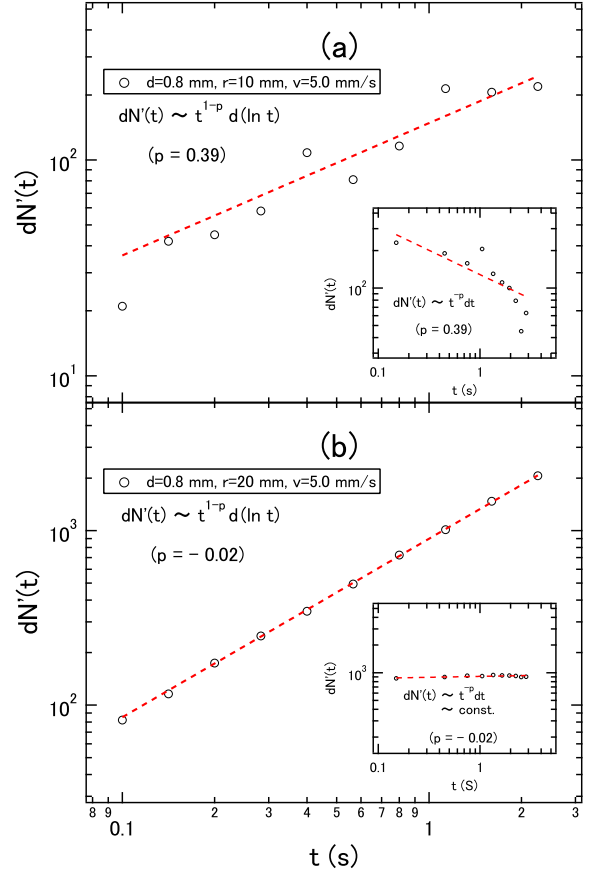}
\end{center}
\caption{(Color online) Power-law histograms with logarithmic bins of the event-occurrence density decay for (a) a nonsteady (OU-like) AE event and (b) a steady (constant) AE event. Since the main plots show the logarithmically binned data, OU law is expressed as $dN'(t) =t^{1-p}d \ln t$. Here, the range $dt=0.1(\sqrt{2})^0 - 0.1(\sqrt{2})^9$~s is used. The least-squares method is used for determining the $p$ value. The insets show the power-law histograms $dN'(t)\sim S(t)dt$ with the constant linear binning $dt=0.3$~s}
\label{fig:Omori}
\end{figure}

However, the current experimental result shows various behaviors. Some $dN'(t)$ distributions follow power-law-like decay (inset of Fig.~\ref{fig:Omori}(a)), others seem to be almost constant (inset of Fig.~\ref{fig:Omori}(b)). The former and latter correspond to the nonsteady and steady processes, respectively. The nonsteady process means the existence of the main event followed by power-law-like decay of subsequent aftershock-like AE events. In other words, the meaningful main event cannot be identified in the latter (steady) case. Actually, even in the power-law (nonsteady) case, the scaled range is not very wide (only one order of magnitude) as can be seen in the inset of Fig.~\ref{fig:Omori}. This implies that it is not very easy to confirm a clear OU law (power-law decay of aftershocks) in the granular AE events. Although the decay range is limited, the inset of Figs.~\ref{fig:Omori}(a) and (b) are significantly different. To discriminate these two phases and discuss the statistical property from the aspect of Markovianity, here we assume the power-law form for $dN'(t)$. The estimated power-law exponent values will play a crucial role to characterize the AE event statistics.

In contrast to calm time distributions that show clear power-law behavior, care must be taken in precisely determining the fitted $p$ value for event-occurrence density distributions. Because most of fitted $p$ values seem less than $1$, the method of MLE cannot be used due to the divergence of Riemann zeta function. Instead, logarithmically increasing bins are employed here. Unequal bin widths are usually used to obtain a more homogeneous number of data per bin than that with a constant bin width, which can reduce statistical errors in the power-law tail due to the poor number of samples~\cite{Newman2005, Sims2007}. Here, the bin widths are given by $0.1(\sqrt{2})^n$~s (e.g., $0.1, 0.1\sqrt{2}, 0.2$ ...~s). Figures~\ref{fig:Omori}(a) and (b) show the histogram of the number of aftershock-like events $dN'(t)$ after the main event obtained by using the logarithmic bins. By applying the logarithmic binning, the power-law exponent of the data plot varies as, 
\begin{equation}
\frac{dN'(t)}{d\ln t} \sim S(t)t \sim t^{1-p}.
\label{eq:log_bin}
\end{equation}
Namely, the slopes in the main plots of Fig.~\ref{fig:Omori} correspond to $1-p$. Although the data considerably scatter in Fig.~\ref{fig:Omori}(a), we assume the power-law behavior at least in the range of $0.1 < t < 3$ s. Then, the Markovianity of the event series after the main event can be evaluated as discussed later in Sec.~\ref{sec:MarkovScaling}. Using this procedure, all the data are fitted by power-law forms. The fitted $p$ values for Figs.~\ref{fig:Omori}(a) and (b) are $0.39$ and $-0.02$, respectively. The corresponding slopes are also shown in the insets as well.

\begin{figure}[!b]
\begin{center}
\includegraphics[width=86mm]{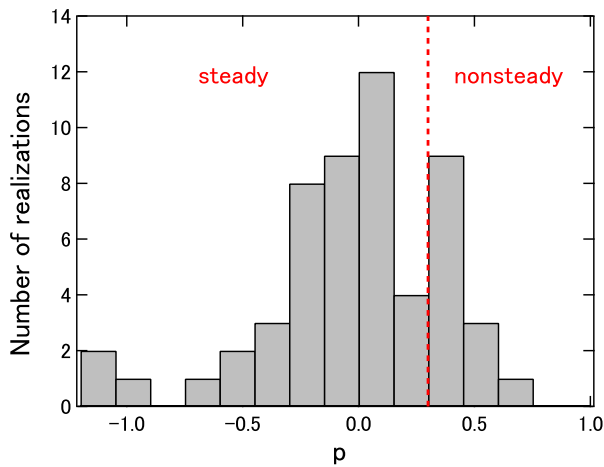}
\caption{(Color online) Histogram of the fitted $p$ value showing a bimodal structure. To distinguish the nonsteady (OU-like) distributions from the steady distributions, the value at the valley between two peaks ($p=0.3$) is used (vertical broken red line).}
\label{fig:p_histogram}
\end{center}
\end{figure}

In Fig.~\ref{fig:p_histogram}, the histogram of the fitted $p$ values for all the dataset is presented. As can be seen in Fig.~\ref{fig:p_histogram}, the histogram shows a bimodal distribution and the valley around $p=0.3$ seems to discriminate two phases. Thus, we define the case of $p>0.3$ as nonsteady (OU-like) states. One can also confirm a small portion of population in the negative $p$ regime. This might be interesting behavior, possibly indicating the precursor of the large event. However, here we only focus on the power-law {\it decay} corresponding to the OU-like state.

\begin{figure}[!b]
\begin{center}
\includegraphics[width=86mm]{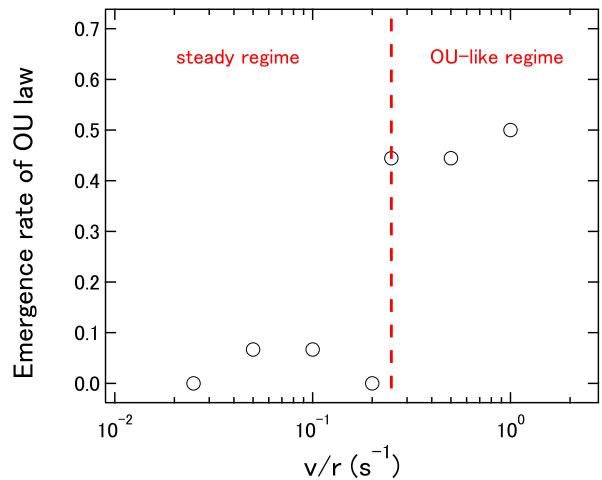}
\caption{(Color online) Effective shear strain rate ($v/r$) dependence of the emergence rate of OU-like behavior. The larger shear strain rate ($v/r\geq0.25$~s$^{-1}$) tends to cause more OU-like behaviors than the small shear strain rate ($v/r< 0.25$~s$^{-1}$). We express these two distinctive regimes as {\it steady} and {\it OU-like} ({\it nonsteady}) regimes.}
\label{fig:Omori_emergence}
\end{center}
\end{figure}

As discussed thus far, some experiments show the OU-like behavior and others do not. What is the most important parameter determining the behavior of the event-occurrence density? To answer this question, we study some parameter dependencies and find that the effective shear strain rate $v/r$ (the penetration speed divided by the radius of a penetrator)~\cite{Katsuragi2015} is relevant to characterize the state. Specifically, the emergence rate of OU-like states is shown as a function of $v/r$ in Fig.~\ref{fig:Omori_emergence}. The emergence rate of OU law is defined by the ratio of the experimental realizations showing $p>0.3$ to the total experimental realizations with the identical $v/r$. Since the number of experimental realizations with the identical $v/r$ is not constant, we employ the normalized occurrence ratio to estimate the emergence frequency. One can confirm that the emergence rate of OU law (nonsteady state) abruptly increases in $v/r \geq 0.25$~s$^{-1}$. Although the complete reproducibility of OU-like behavior is not established even in the range of $v/r\geq0.25$~s$^{-1}$, Fig.~\ref{fig:Omori_emergence} implies that $v/r=0.25$~s$^{-1}$ is the marginal value between the steady and nonsteady (OU-like) regime.

\begin{figure}[!b]
\begin{center}
\includegraphics[width=86mm]{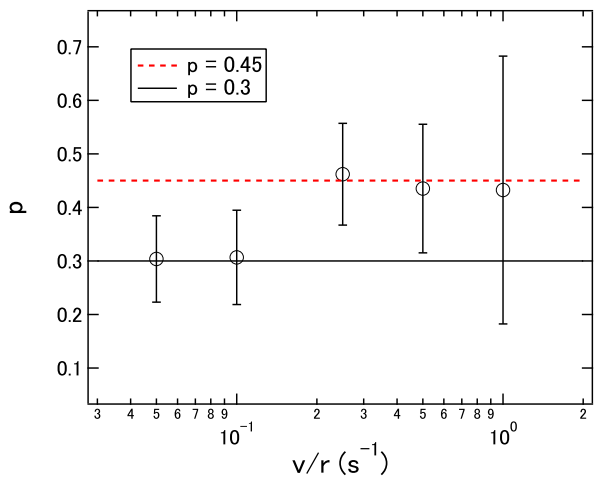}
\caption{(Color online) Effective shear strain rate ($v/r$) dependence of the fitted $p$ value. The broken red line represents the level of $p=0.45$ while the solid black line refers to the marginal value $p=0.3$. The $p$ value obtained in OU-like regime ($v/r\geq0.25$~s$^{-1}$) is almost independent of $v/r$.}
\label{fig:p_vs_vr}
\end{center}
\end{figure}

Next, to verify the $v/r$ dependence of $p$, the averaged $p$ value of OU law is plotted as a function of $v/r$ in Fig.~\ref{fig:p_vs_vr}. In the OU-like regime ($v/r\geq0.25$~s$^{-1}$), the $p$ values show a roughly constant value, $p\simeq 0.45$, in contrast to the steady regime suggesting $p\simeq 0.3$, which might result from the defined marginal value between steady regime and OU-like regime ($p=0.3$). Since the emergence rate of OU-like behavior is very low in $v/r< 0.25$~s$^{-1}$, the value is sensitive to the threshold. However, the data errors in Fig.~\ref{fig:p_vs_vr} are considerably large, thus the difference between $p=0.3$ and $p=0.45$ is not very clear.

\subsection{Markov scaling}
\label{sec:MarkovScaling}
Using two exponents $\mu$ and $p$ defined in Eqs.~(\ref{eq:P_tau}) and (\ref{eq:S_t}), Markovianity of the event time series after the main event can be discussed in terms of a scaling law. The specific values used in the evaluation of the Markov scaling are $\mu$ (Fig.~\ref{fig:calm_time_distribution}) and $p$ (Fig.~\ref{fig:p_vs_vr}) computed from the data after the main event. The scaling law used here was originally developed and applied to real seismicity. As a result, non-Markov nature of earthquake aftershocks was reported~\cite{Abe2009,Abe2012}. Here, we slightly expand this scaling law. If a process of event occurrence is Markovian, the following equation holds~\cite{Bardou2002}:
\begin{equation}
S(t)=P(t)+\int_0^t P(t-t')S(t') dt',
\label{eq:Kolmogorov}
\end{equation}
which can be derived from the Kolmogorov forward equation~\cite{Barndorff2000}. Then, the Laplace transformation of Eq.~(\ref{eq:Kolmogorov}) yields
\begin{equation}
\mathcal{L}[S](s)=\frac{\mathcal{L}[P](s)}{1-\mathcal{L}[P](s)},
\label{eq:Lap_Kormogorv}
\end{equation}
where $\mathcal{L}[f](s) = \int_0^{\infty}e^{-st}f(t)dt$. Here, we assume that both $P(t)$ and $S(t)$ decay following the power-law forms as written in Eqs.~(\ref{eq:P_tau}) and (\ref{eq:S_t}) 
for a large value of $t$. Then, the Laplace transformations of $P(t)$ and $S(t)$ result in different expressions depending on the ranges of the exponents. If the exponents $\mu$ and $p$ are in the ranges
\begin{equation}
0<\mu<1 \mbox{ and } 0<p<1,
\label{eq:mu_p_range_1}
\end{equation}
the Laplace transformations of $P(t)$ and $S(t)$ behave as
\begin{equation}
\mathcal{L}[P](s)\sim 1-\alpha s^\mu,
\label{eq:Lap_mu}
\end{equation}
\begin{equation}
\mathcal{L}[S](s)\sim\frac{1}{s^{1-p}},
\label{eq:Lap_p}
\end{equation}
for a small limit of $s$, where $\alpha$ is a positive constant. Substituting Eqs.~(\ref{eq:Lap_mu}) and (\ref{eq:Lap_p}) into Eq.~(\ref{eq:Lap_Kormogorv}), we obtain a simple scaling relation~\cite{Abe2009,Abe2012}:
\begin{equation}
p+\mu = 1.
\label{eq:p_mu_1}
\end{equation}
However, if the exponent $\mu$ lies in the range
\begin{equation}
1<\mu<2,
\label{eq:mu_range_2}
\end{equation}
the resultant Laplace transformation becomes
\begin{equation}
\mathcal{L}[P](s)\sim1-\beta s, 
\label{eq:Lap_mu_2}
\end{equation}
for a small limit of $s$, where $\beta$ is a positive constant. Using Eqs.~(\ref{eq:Lap_p}) and (\ref{eq:Lap_mu_2}), Eq.~(\ref{eq:Lap_Kormogorv}) results in
\begin{equation}
p = 0.
\label{eq:p_zero}
\end{equation}
Note that this result is inconsistent with the assumed range of Eq.~(\ref{eq:mu_p_range_1}); this implies that Eq.~(\ref{eq:p_zero}) is only an asymptotic solution. In the case of $\mu>2$, the scaling relation fulfilled in a Markov process does not exist. 

\begin{figure}[!b]
\begin{center}
\includegraphics[width=86mm]{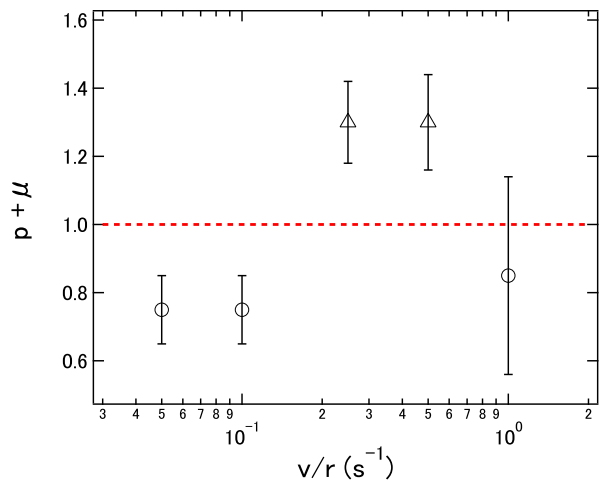}
\caption{(Color online) $p+\mu$ as a function of the effective shear strain rate $v/r$. The broken red line indicates the level $p+\mu=1$ (Markov scaling law). Circles distribute around $p+\mu\simeq 1$ (or $p+\mu \lesssim 1$) while triangles clearly beyond the value.}
\label{fig:p_mu}
\end{center}
\end{figure}

Equation~(\ref{eq:p_mu_1}) indicates the criterion for the Markovianity in the OU-like (nonsteady) event time series in the range of Eq.~(\ref{eq:mu_p_range_1}). To verify Markovianity of the granular aftershock-like AE events showing OU-like behavior, $p+\mu$ as a function of $v/r$ is depicted in Fig.~\ref{fig:p_mu}. The broken red line in Fig.~\ref{fig:p_mu} indicates the Markov scaling law (Eq.~(\ref{eq:p_mu_1})). One can confirm that some data (triangles) show clearly large values ($p+\mu > 1$) that are somewhat similar to earthquake aftershocks~\cite{Abe2009,Abe2012}. In the current experimental result, however, the other data points (circles) distribute around $p+\mu \simeq 1$ (or $p+\mu \lesssim 1$). This means that there might be a certain parameter range in which the event time series can be regarded as a Markov process. This result is contrastive to the real seismic activity that always shows $p+\mu>1$, i.e., a non-Markov property. 

\begin{figure}[!b]
\begin{center}
\includegraphics[width=86mm]{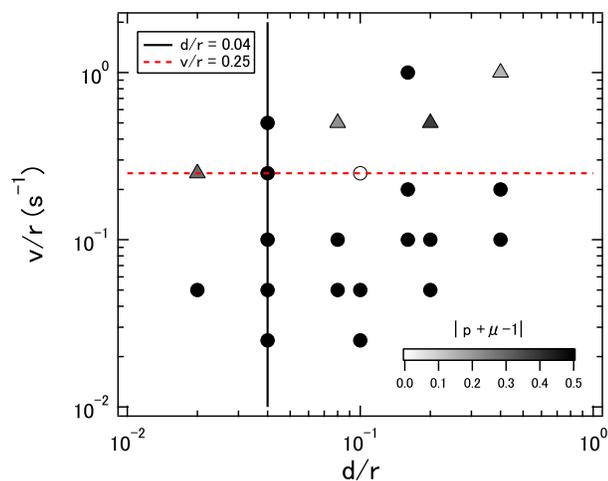}
\caption{(Color online) Phase diagram of the temporal statistics of granular AE events. x- and y-axes represent $d/r$ and $v/r$, respectively. The meaning of each symbol is as follows. The triangles represent the cases in which nonsteady (OU-like) behaviors can be observed. Note that triangles does not mean the complete reproducibility of OU-like behavior. As mentioned in the text, three experimental realizations for each experimental conditions were carried out. When two of three realizations show OU-like behavior, the triangle mark is used. The gray scale in triangles indicates the value of $|p+\mu -1|$ as shown in the legend. The open circle indicates the steady Markov process ($1<\mu<2$, $p \simeq 0$), while the filled circles indicate the steady non-Markov process ($0<\mu<1$ or $2<\mu$, $p \simeq 0$). The vertical line $d/r=0.04$ is the boundary between brittle-like and plastic-like regimes~\cite{Matsuyama2014}. The horizontal dashed line $v/r=0.25$~s$^{-1}$ is considered to be the characteristic shear strain rate above which the OU-like behavior can often be observed.}
\label{fig:p_mu_diagram}
\end{center}
\end{figure}

Actually, the effective shear strain rate $v/r$ is not a unique parameter to characterize the system. To describe the global behavior of all granular AE events in this study, the normalized grain size $d/r$ is also used here. Indeed, $d/r$ can be a characteristic parameter to classify the event-size distribution (GR law)~\cite{Matsuyama2014}. $d/r=0.04$ is regarded as a marginal value between brittle-like and plastic-like behavior. As discussed in the last subsection, on the other hand, $v/r=0.25$~s$^{-1}$ is considered as a marginal value between the steady and nonsteady (OU-like) states. Therefore, we make a phase diagram as shown in Fig.~\ref{fig:p_mu_diagram}, where the vertical and horizontal axes indicate the effective shear strain rate $v/r$ and normalized grain size $d/r$, respectively. Although it is difficult to draw clear phase boundaries, there might be several classes of behavior in the phase diagram. In the large $v/r$ and large $d/r$ regime, the nonsteady (OU-like) event time series, which is shown by triangles, can be frequently observed. In this OU-like regime, the value of $|p+\mu-1|$ is indicated by the gray scale in each triangles as long as $\mu$ is less than $1$~(i.e., the range defined by Eq.~(\ref{eq:mu_p_range_1})). Namely, the dark triangle means that the process is non-Markov. The steady ($p \simeq 0$) cases are represented by circles. The open and filled circles correspond to the steady Markov ($1<\mu<2$, $p \simeq 0$) and steady non-Markov ($0<\mu<1$ or $2<\mu$, $p \simeq 0$), respectively.

\section{Natural time analysis}
\label{sec:sec4}
\subsection{Variance $\kappa_1$}
\begin{figure}[!b]
\begin{center}
\includegraphics[width=86mm]{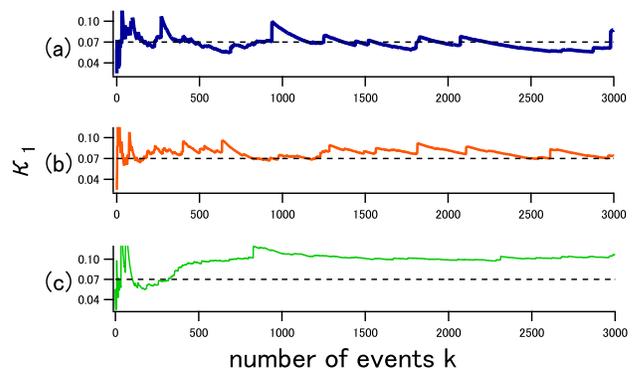}
\caption{(Color online) Evolution of $\kappa_1$ as a function of the number of events $k$ for various experimental conditions of (a) $d=2.0$, (b) $0.8$, and (c) $0.4$~mm from the top to the bottom, respectively. The other parameters are fixed as $r=10$~mm and $v=5.0$~mm~s$^{-1}$. The evolutions of each $\kappa_1$ are shown up to $k=3000$. The horizontal broken black lines represent $\kappa_1=0.07$ indicating the criticality of the system. The top (a) and middle (b) data look fluctuating around $0.07$ while the data in (c) is clearly offset.}
\label{fig:kappa_chi}
\end{center}
\end{figure}

\begin{figure}[!b]
\begin{center}
\includegraphics[width=86mm]{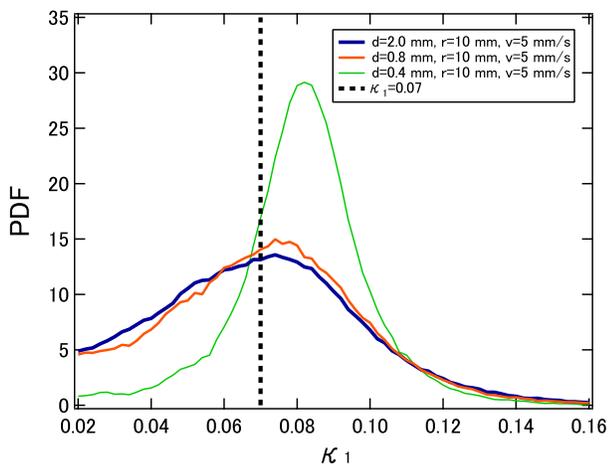}
\caption{(Color online) PDF of the $\kappa_1$ value in the experimental conditions same as Fig.~\ref{fig:kappa_chi}. The color code is identical to that in Fig.~\ref{fig:kappa_chi}. The vertical broken line indicates $\kappa_1=0.07$. The peak values of the blue ($d=2.0$~mm) and red ($d=0.8$~mm) curves are around $0.07$.}
\label{fig:kappa_PDF}
\end{center}
\end{figure}

Next, the natural time analysis is applied to the identical dataset. In the definition of natural time, calm time is completely neglected. Instead, the order and magnitude of events are utilized to characterize the event statistics. By simply applying Eq.~(\ref{eq:kappa_def}) to the dataset, we can readily calculate $\kappa_1$. Here we use the squared maximum amplitude of each AE event for its released energy value $Q_k$~\cite{Matsuyama2014}. Some examples of $\kappa_1(k)$ are shown in Fig.~\ref{fig:kappa_chi}. As expected, $\kappa_1$ looks fluctuating around $\kappa_1 \simeq 0.07$ in some data, which indicates the criticality of the system. However, this tendency is not very universal. For instance, the asymptotic value of $\kappa_1(k)$ in Fig.~\ref{fig:kappa_chi}(c) seems to be different from $0.07$. To verify the $\kappa_1$ behavior in more detail, the probability density function (PDF) of $\kappa_1$ is computed by the following procedure~\cite{Varotsos2005, Varotsos2006}. First, the $\kappa_1$ value is computed from the natural time windows for $6-40$ consecutive events. Second, this process is performed for all events by scanning the whole data set. Figure~\ref{fig:kappa_PDF} shows the PDF computed from the data used in Fig.~\ref{fig:kappa_chi}. The most probable value $\kappa_{1,p}$ estimated by the peak location of the PDF depends on experimental conditions. Particularly, the grain diameter $d$ seems to be an important parameter. This $d$-dependent tendency is similar to the behavior of $\gamma$ in Eq. (\ref{eq:size_power_law}) (event-size distribution exponent)~\cite{Matsuyama2014}.

\subsection{Returns distribution}
Although the measurement of $\kappa_1$ is easy and useful to briefly characterize the critical state in the time series, it is in general not sufficient to evaluate the criticality of the event time series. Caruso $et~al.$~\cite{Caruso2007} have suggested another way for characterizing the criticality in the system. 

\begin{figure}[!b]
\begin{center}
\includegraphics[width=86mm]{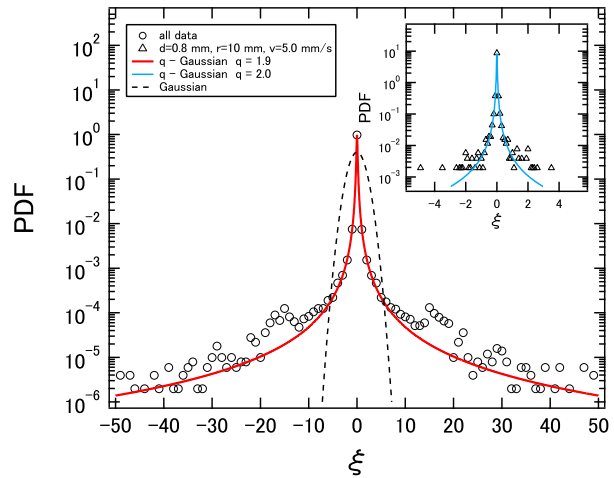}
\caption{(Color online) PDF of the normalized returns $\xi$ computed from the whole AE events in all experimental conditions. The data obtained in the current experiment is better fitted by $q$-Gaussian (solid red curve) than normal Gaussian (broken black curve). The inset shows an example of single PDF in the experimental conditions of $d=0.8$~mm, $r=10$~mm, and $v=5.0$~mm~s$^{-1}$.}
\label{fig:PDF_q_gauss}
\end{center}
\end{figure}

The variable {\it returns} $x(k)$ are defined as $x(k)=Q_{k+1}-Q_{k}$. Namely, $x(k)$ corresponds to the energy difference between two successive events. Here the returns are normalized to the mean $\langle x\rangle$ and standard deviation $\sigma$, $\xi=(x-\langle x\rangle)/\sigma$. Then the PDF of $\xi$ is calculated for each experimental condition (e.g., the inset of Fig.~\ref{fig:PDF_q_gauss}). As a result, it is clarified that the functional form of the PDF is independent of experimental conditions. Therefore, to obtain better statistics, the whole data of the normalized returns under various experimental conditions are merged into a single PDF. The entire PDF as a function of $\xi$ is shown in the main plot of Fig.~\ref{fig:PDF_q_gauss}, in which the PDF has much broader tails than normal Gaussian distribution $f(\xi)=\exp(-\xi^2/2)/\sqrt{2\pi}$ (broken black curve). Instead, q-Gaussian form $f(\xi)=A_q[1-(1-q)\xi^2/B_q]^{1/(1-q)}$ can fit the data well (red curve), where $A_q$ and $B_q$ are constants~\cite{Tsallis2005}. The computed $q$ value is $q=1.9$. If the data of a single experimental run are used for the PDF as shown in the inset of Fig.~\ref{fig:PDF_q_gauss}, the agreement between $q$-Gaussian and data is limited particularly in the tail part. However, the qualitative tendency of the distribution is basically identical. Because the $q$-Gaussian PDF of $\xi$ can be observed in the SOC model satisfying both the power-law event-size distribution and the finite size scaling~\cite{Caruso2007}, this result provides the supportive evidence for the presence of the critical state.

\subsection{Relation between $\kappa_1$ and $\gamma$}
\begin{figure}[!b]
\begin{center}
\includegraphics[width=86mm]{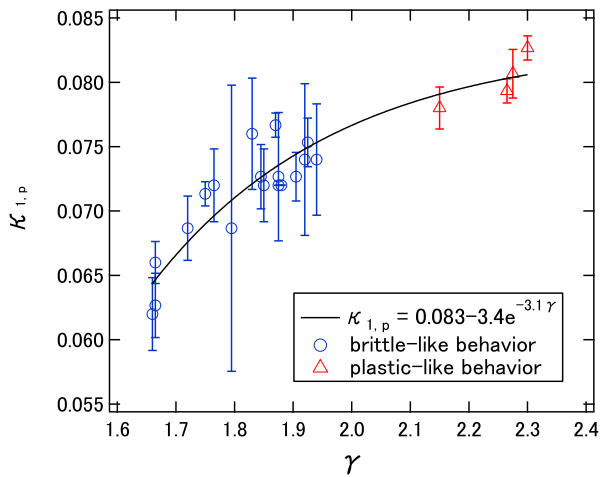}
\caption{(Color online) The most probable value $\kappa_{1,p}$ located by the peak of the PDF as a function of the power-law event-size distribution (GR law) exponent $\gamma$. The colors and symbols represent the difference between brittle-like regime (blue circles) and plastic-like regime (red triangles). The blue circles tend to distribute around $0.07$. The asymptotic value estimated by the fitting is approximately $0.083$, which coincidentally corresponds to $\kappa_1$ of a {\it uniform} distribution~\cite{Varotsos2005(1),Varotsos2003,Varotsos2003(1),Varotsos2004}.}
\label{fig:kappa_gamma}
\end{center}
\end{figure}

Finally, we consider the relation between natural time analyses and event-size distributions. As shown in Figs.~\ref{fig:kappa_chi} and \ref{fig:kappa_PDF}, the $\kappa_1$ value is mainly affected by the grain diameter $d$. Additionally, the exponent of power-law event-size distributions $\gamma$ (Eq.~\ref{eq:size_power_law}) also depends on $d$~\cite{Matsuyama2014}. Thus, $\kappa_1$ and $\gamma$ might show a certain relation. To check the relationship, $\kappa_1$ vs.~$\gamma$ is plotted in Fig.~\ref{fig:kappa_gamma}. As expected, a correlation between them can be confirmed. The symbols (or colors) in Fig.~\ref{fig:kappa_gamma} indicate the difference of deformation mode: brittle-like mode characterized by the smaller $\gamma$ (blue circles) and plastic-like mode characterized by larger $\gamma$ (red triangles). The solid black curve in Fig.~\ref{fig:kappa_gamma} provides a fitting by exponentially asymptotic function $\kappa_{1,p}=0.083-3.4e^{-3.1\gamma}$. The asymptotic value coincidentally agrees with $\kappa_u=0.083~(\simeq 1/12)$ of a {\it uniform} distribution~\cite{Varotsos2005(1),Varotsos2003,Varotsos2003(1),Varotsos2004}. In Fig.~\ref{fig:kappa_gamma}, $\kappa_1=0.07$ seems to be satisfied in the experimental conditions of brittle-like behavior. Put simply, the brittle-like regime shows $\kappa_1 \simeq 0.07$ while $\kappa_1$ approaches $0.083$ by increasing $\gamma$, i.e., in more plastic-like (flowing) regime. The condition that $\kappa_1$ approaches 0.083 means that the critical state is not established in the time series of AE events.  

\section{Discussion}
\label{sec:sec5}
Thus far, various temporal analysis methods have been applied to the granular AE event data. Here let us discuss their physical meaning and relations.

In the actual time analysis, power-law exponents for the calm time distribution and the decay of the event-occurrence density were measured. All of the experimental data show power-law forms for the calm time distribution. However, the obtained exponent value $\mu$ significantly depends on the penetration speed $v$. This tendency is in contrast to the case of power-law event-size distributions (GR law), in which the exponent $\gamma$ is mainly determined by the grain size $d$ and almost independent of the penetration speed $v$. For the event-occurrence density, power-law behavior is not universal. The emergence rate of the power-law decay (OU-like behavior) can be characterized by the effective shear strain rate $v/r$. Since the dimension of $v/r$ corresponds to the inverse of time, this quantity indirectly represents a characteristic timescale in the penetration system. Additionally, the time range of OU-like behavior is equal to or less than the order of $10^0$~s (e.g., Fig.~\ref{fig:Omori}), suggesting the relaxation process occurs within the characteristic timescale ($0.25^{-1}=4$ s). Therefore, it is natural that the temporal property such as the event-occurrence density can be sorted by $v/r$. In the range of the OU-like regime in the granular AE events ($v/r\geq0.25$~s$^{-1}$), the power-law exponent $p$ shows an almost constant value $p \simeq 0.45$ independent of $v/r$ (Fig.~\ref{fig:p_vs_vr}). This $p$ value is less than the typical value for real seismicity and AE events from the microfracturing of rocks, $p \simeq 1$~\cite{Ojala2004}. The reason for this discrepancy remains unsolved. Perhaps, this difference originates from the peculiar nature of deformation in bulk granular matter.

For the natural time analysis, on the other hand, the variance $\kappa_1$ is related to the exponent of power-law event-size distributions (GR law) $\gamma$. Since the exponent $\gamma$ is mainly determined by the grain size $d$~\cite{Matsuyama2014}, the $\kappa_1$ in granular AE event time series is also related to the grain size $d$. Additionally, in the natural time analysis, the interval time between events (calm time) is completely neglected and only the order and amplitude of events are used. This is the reason why the geometrical parameter such as $d$ or $d/r$ becomes more essential than the temporal parameter $v$ or $v/r$ in the natural-time-related analysis. 
These two analysis methods (actual time and natural time analyses) are complementary to each other.

As a matter of fact, the correlation between $\kappa_{1,p}$ and $\gamma$ similar to Fig.~\ref{fig:kappa_gamma} has also been observed in the artificially randomized (shuffled~\cite{Varotsos2004}) event series data~\cite{Varotsos2006}. However, the agreement remains qualitative; the specific value ranges are slightly different between the randomized event series and the current experimental data. This difference might come from the effect of memory among events as the randomized data do not have memory. 

The statistical behavior of granular AE events becomes similar to that of real seismicity when both $d/r$ and $v/r$ are in the appropriate ranges: $d/r > 0.04$ and $0.25<v/r<1.0$~s$^{-1}$ (Figs.~\ref{fig:p_mu}, \ref{fig:p_mu_diagram}, and \ref{fig:kappa_gamma} and Ref.~\cite{Matsuyama2014}). In this regime, the values of $\gamma$, $\kappa_1$, and $p+\mu$ show similar values between the granular AE events and real seismic activity. Note that, however, the specific values of $p$ and $\mu$ are different between the granular AE events and real seismicity. Furthermore, this coincidence in characteristic quantities might not directly mean the correspondence of underlying physical mechanisms. 

For instance, OU law for real seismic activity can be considered as the relaxation process after the mainshock. When the penetration speed is rapid, the available time for the relaxation becomes relatively short (might be insufficient) in general. Thus, it is difficult to see the relaxation in the large $v/r$ regime. On the other hand, the large $v/r$ is necessary to reproduce OU-like behavior in the granular AE experiment. This implies the large $v/r$ might result in the large shear field which causes a number of aftershock-like AE events. These two effects will compete with each other and the qualitative tendency seems to be opposite. As a result, the complex statistical properties are observed. To simplify the problem, the controlled experiment in a different geometrical setup should be performed. The current experimental setup actually originates from the previous researches concerning the slow-penetration drag force in granular matter~\cite{Katsuragi2012a,Katsuragi2012b,Matsuyama2014}. The simpler setup such as a simple shear, which can mimic the fault slip, might be better for future studies. 

\section{Conclusion}
\label{sec:sec6}
The statistical property of granular AE events was investigated using two approaches: actual time analysis (Sec.~\ref{sec:sec3}) and natural time analysis (Sec.~\ref{sec:sec4}). In the actual time analysis, the calm time distribution always shows a power-law form while OU-like behavior can only be frequently observed in the range of $v/r\geq0.25$~s$^{-1}$. In the OU-like (nonsteady) regime, non-Markov behavior is observed in a particular $v/r$ regime. However, the steady (not OU-like) behavior is actually dominant in the granular AE events. To control the emergence of OU-like behavior and Markovianity, the appropriate tuning of $v/r$ (inverse of the timescale) is necessary. Natural time analyses revealed that the $\kappa_1$ value distributes around $0.07 - 0.083$. In addition, $q$-Gaussian fits the distribution of the returns (i.e., energy difference among events). This result supports the SOC state of the system. The $\kappa_1$ value can be related to the exponent of power-law event-size distributions (GR law) $\gamma$. This means that the grain size $d$ is the important parameter for $\kappa_1$ because $\gamma$ is directly related to $d$~\cite{Matsuyama2014}. We find that $\kappa_1 \simeq 0.07$ can be established in the brittle-like regime ($d/r>0.04$).

In summary, statistical properties of seismic activity can be mimicked by the granular AE events in the range $d/r > 0.04$ and $0.25 < v/r< 1.0$~s$^{-1}$. Although the current experimental system is different from the microfracturing of rocks and geological scale phenomena, the AE data obtained from a plunged granular matter exhibits some similarities with geological scale phenomena like earthquakes in terms of actual time and natural time analyses.


\section*{Acknowledgments}
We would like to appreciate S. Abe, T. Hatano, S. Watanabe, H. Kumagai, S. Sirono, and T. Morota for fruitful comments and discussions. Besides, we also thank K. Matsuyama for taking the data used in this study. 


\bibliography{TMN}

\end{document}